# Scaling in polymers: I. The *ortho*-fused spiral-benzenes.

H. G. Miller[1,2], C. J. H. Schutte[1,2] and P. H. van Rooyen[1]

[1] Department of Chemistry, University of Pretoria - 0002 Pretoria, South Africa
[2] Centre for Advanced Studies, University of Pretoria 0002, Pretoria, South Africa

## Abstract

Analogous to a model that predicts the linear scaling of the binding energy of a nucleus from the number of nucleons, a simple model was developed to account for the observed linear variation of the quantum-chemically computed total electronic energy of the fully-optimized structures of a homologous series of polymers. This model was tested with both *ab-initio* DFT and molecular mechanics methods on the *ortho*-fused *spiral*-benzenes. Both methods predict linear scaling of total polymer energy with increasing number of repeating units added. Since this is also the case for the linear *ortho*-fused zigzag-benzenes and other polymers, it is postulated that the model is applicable to polymers in general. It may, therefore, be used to predict physical properties of long-chain polymers.

## Introduction

In a series of papers by Hausser *et al* [1-4] the relation between the absorption of light and the number $n$ of double bonds in a molecule was investigated [1-4]. Absorption measurements made on solutions of the diphenylpolyenes $C_6H_5$-(CH=CH)$_n$-$C_6H_5$ with $n$ = 1 to 7 indicated that both the height of the absorption band and the wave length of the absorption band maximum are nearly linear functions of $n$ [1]. A similar relationship was found for a series of polyene aldehydes and polyene carboxylic acids, $CH_3$-(CH=CH)$_n$-CHO with $n$ = 1 to 3, and $CH_3$-(CH=CH)$_n$-COOH with $n$ = 1 to 4 [2].

Some properties of polymers, therefore, are predetermined by both the chemical and physical characteristics of their respective constituent repeating units themselves and the (gross) characteristics of the polymer itself, such as the number of repeating units in the molecule. Two effects must be distinguished here, namely

(i) *quantum mechanical effects* caused by changes in the nature of chemical bonds between the repeating units when another unit is added (in this is included what the chemists call delocalization energy that occurs whenever double bonds are conjugated, as well as the individual energies of the added atoms), and

(ii) changes in weak *interatomic forces* between atoms or repeating units not bonded to one-another (such as steric hindrance between bulky atoms and groups, Van der Waals repulsive effects, etc.).

These two effects could, respectively, be classified as caused by *short-range forces* and *long-range forces*. From the measurements of Hausser *et al*. it seems that the *sum total of both these effects* saturates very quickly for some physical properties of polymers for the addition of each further repeating unit. It is therefore not unreasonable to expect that the effect of long-range forces is dominated by that of short-range forces.

A somewhat similar situation occurs in nuclei where the individual nucleons are considered to retain their structure, and the nuclear force, although more complicated in nature, is short-ranged and dominates the weaker long-ranged Coulomb interaction. In the nuclear case, this led to a simple semi-empirical formula for the binding energy of a nucleus which contains $A$ nucleons [5-7] which is a function of $A$, and $Z$ (the number of protons in nucleus). Because the nuclear force is short-ranged, it quickly saturates and the binding energy per nucleon becomes constant and an extrapolation to the bulk (nuclear matter) is possible [7]. However, if polymer structure is dependent on the sum of the nearest neighbor interactions and the long-ranged interactions are negligible, then a similar extrapolation must be possible for polymers.

A *quantum-chemical ab-initio calculation* of the total molecular energy $E$ at the ground state of the potential- minimum configuration of a molecule, called the fully-optimized geometrical structure, yields the sum total of the two effects discussed above. In particular, a systematic calculation of the ground state energy $E_n$ of a polymer made of $n$ repeating units, $E_n$, should thus display the following behavior:

*For sufficiently large value of n the successive differences $\Delta E = E_n - E_{n+1}$ should become constant.*

If this can be computationally shown to be true by *quantum-chemical methods*, it would be of significant value, as it would enable the confident extrapolation of the physical properties of a polymer to large values of $n$, (i.e. to values for large polymers) well beyond the current computational limits of even the largest computers.

A *Molecular Mechanics* (MM) *calculation* is based upon a set of empirically-determined harmonic force constants that operate between bonded and non-bonded atoms to find the minimum-energy geometric structure of a molecule. Although an MM calculation does not formally include electronic forces directly in the force field, they are inherently present, since they also contribute to the harmonic potential function from which the (empirical) force constants are derived. The same argument, therefore, holds for the energy values obtained from MM calculations, and for a sufficiently large values of $n$ the successive differences $\Delta E = E_n - E_{n+1}$ should also become constant.

The model

Consider a polymer made up of $n$ repeating units. The results of Hausser et al.*[1-4]* between some physical properties of successive polymers strongly suggest that the ground state energy of a polymer, $E(n)$, should *linearly* scale like $n$, that is

$$E_n = kn \qquad\qquad (1)$$

where $k$ is a constant. Equation (1) is consistent with a simple Fermi Gas model estimate [8].

Consider $N$ electrons (the number of electrons in a molecule), each of which has a single particle energy given by

$$E^* = p^2/2m^* \qquad\qquad (2)$$

where the effects of the interactions are taken into account by replacing the bare mass of the electron, $m$, by its effective mass, $m_*$. It follows that [8]

$$E_N / N = (3/5)\varepsilon^*_f \qquad\qquad (3)$$

where $\varepsilon^*_f = p^2_f/2m_*$ is the Fermi energy. If the density, $\rho = N/V$ remains roughly constant, as $N$ grows (i.e. as the number of molecules increases), the Fermi momentum, $p_f$, (which varies as $\rho^{1/3}$) also remains constant. Furthermore, if the renormalization effects are not large ($m^* \approx m$), which is not unreasonable in the case of molecular physics, or are only dependent on the density, the effective mass will not have a strong dependence on $N$. Hence $\varepsilon_f$ remains constant, and the Fermi gas model yields a result which is consistent with equation (1).

In order to test the validity of these general considerations, we employ two different quantum-chemical computational methods to establish a correlation between the increase of total molecular energy (including conformational and steric energy) as a function of the number of repeating units added when building polymers of increasing size up to $n = 10$. To illustrate this, the optimized total energies $E_n$ of the first ten members a homologous series aromatic polymer with an increasing number $n$ of *ortho*-fused aromatic six-membered rings, which exemplify strong non-bonded (long-range) repulsive interactions, have been determined, using both *molecular mechanics* and *quantum mechanical methods*.

The homologous series of *ortho*-fused aromatic ring polymeric molecules, called the *spiral-benzenes* in this paper, was investigated. Figure 1 shows the chemical diagrams of the first few members, while Figure 2 displays a 3D-view of the spiral geometry of a typical such molecule). For the sake of *internal* consistency, these molecules are referred-to by simply *prefixing* the number, $n$, the number of formal benzene hexagons in the structural formula to the term *spiro*benzene, that is, 1-*spiro*benzene (benzene), 2-*spiro*benzene (naphthalene), 3-*spiro*benzene (phenanthrene), 4-*spiro*benzene (*benzo*phenanthrene), etc. Fused-ring polycyclic molecules containing only multiple benzene rings are usually planar due to the preferential delocalization of the π-electrons occurring at each C-atom, except when steric and other strain factors occur, which are strong enough to force the molecule out of planarity and to reduce the contribution of the π-electrons to the energy of the molecule. The first three members of this series are planar. There is a rather short non-bonded H4…H5 distance of 2.015 Å in $n = 3$, phenanthrene, which does cause some strain in the molecule, but it is not enough force it out of planarity. In the case of the homologous series of ortho-fused molecules illustrated in Figure 1 the deviations from planarity for $n \geq 4$ and larger, are caused by strong steric hindrance between pairs of non-bonded carbon hydrogen atoms , namely, the pairs (C…C), (C…H) and.(H…H).  In the case of $n = 4$ this repulsive interaction is caused by the strong steric hindrance between the non-bonded atoms H1 and H12 (IUPAC numbering shown in Figure 1); this interaction is reduced by forming the spiral structure, since it separates the two almost-overlapping hydrogen atoms by increasing their distance to 2.335 Å, according to the computational results of the present paper.  However, for $n > 5$, non-bonded (C…C), (C…H) and.(H…H ) interactions contribute to the forces that cause such a molecule to compromise by becoming spiral. These spiral structures may be either left-handed or right-handed, both having identical total molecular energy as calculated by any quantum-chemical method. We report here the results of spiral molecules having a right-handed sense, plotting the quantum-chemically computed total molecular energy $E_n$ for the series of *ortho*-fused spiralbenzenes from $n = 1$ to 10 against $n$, the number of formal benzene rings in the structure to determine the shape of the resulting curve. From the preceding arguments, it is clear that weak (H…H) non-bonded repulsive forces are already introduced in the $n = 3$ molecule, but become stronger in the case of the $n = 4$ molecule, causing it to develop a spiral twist. These distortions due to the non-bonded forces are only fully developed for the C…C, H…H and C…H non-bonded interactions for the n = 5 molecule. It is, hence, expected that from n = 5 step-wise increase in the energy due to increase in the number of benzene rings in the sense of the model proposed above, will saturate and that the curve would become strictly linear from $n = 5$ onwards; for values of n < 5; the curve may show some evidence of non-linear behavior.

Clearly, if linear scaling is indeed observed as the model predicts, an extrapolation to large values of $n$ may be a viable means to reliably estimate the total energy of large polymers which may not be accessible within current computational limits.

Computational details

All *ab-initio quantum-chemical computations* were carried out on an IBM-cluster with a pre-compiled set of Gaussian-03 molecular orbital programs [10] configured for parallel computing under LINUX [11]. Default computational settings for Gaussian-03 *ab-initio* DFT/B3LYP calculations were implemented, while the 6-311G basis set was used as described in the Gaussian manual [12]. This basis set was chosen instead of the more elaborate 6-311++G(3df,3pd) basis set because it represents a balance between computational level and computational economy. More details about the background of the computational methodology can be found in the books by Hirst [13], Kohanoff [14], and Foresman and Frisch [15], as well as in references cited in the Gaussian Manual [12].

The molecular structure of each spiral molecule was first optimized using the molecular symmetry point group $C_1$, although all these spiral molecules actually belong to the molecular point group $C_2$ within tolerances of 0.01 Å. The reason for choosing $C_1$ as the refining point group is that for n > 4, it was found that all spiral molecules with *n* even do not properly refine, but cycle between two potential energy minima ($C_1$ and $C_2$) which are very near to one another. Under point group $C_1$ it was found that all molecules studied refined satisfactorily within the cut-off values of Gaussian 03.

In order to determine whether the optimized geometry of each of the spiral benzenes, where the slope approaches $(\partial E/\partial X) \rightarrow 0$, occurs at a *potential minimum* and not at a saddle point, the harmonic molecular vibrational modes and their associated frequencies were calculated at the optimized geometric configurations [16]. It was established that all optimized geometries of the molecules studied yielded only positive vibrational frequencies, that is, all optimized geometries were thus determined at *potential energy minimum positions*, except for $n = 15$ and $n = 20$, for which the frequency calculations could not be performed on our cluster due to the computational requirements of Gaussian 03 for such large molecules using the fairly large basis set 6-311G. However, the optimization sequences we obtained for these cases proceeded normally, and we assume that the energies we report, were obtained for structures at potential minima.

*Molecular mechanics calculations* were also done with Gaussian-03, using the *universal force field* UFF described the Manual [12]. The relative energy scales of MM-optimizations and DFT *ab-initio* optimizations are different, since MM calculations do not directly incorporate electron interaction energies.

In order to test the *extrapolation* of the linear model described here beyond $n = 10$, the energies of polymers with $n = 15$ and $n = 20$ benzene rings were also calculated, using the identical molecular mechanics and quantum-chemical computational methodologies on different computers. These are then compared with the extrapolated values obtained from a linear fit of the results of $n = 1, 2, \dots, 10$.

With increasing polymer length *n* it was found that the *total molecular energy $E_n$*, as calculated with both molecular mechanics (MM) as well as with *ab initio* methods, scales *linearly with polymer length* as shown in Table 1 and Figures 3 and 4. For the *ab-initio* G03 results (Figure 3) the statistical parameters of the least-squares fitted linear relationship $E = a + kn$ for n >4 were found to be:

$a$ = -78.642213 hartree          slope $k$ = -153.62842 hartree/ring,
$r^2$ = 1.000                              $\sigma$ = 0.0010340.

(Note that $a \neq 0$, since *E* does go to zero linearly for $n < 3$ for the reasons given above; this is addressed in more detail in paper II of the series.) Using the these parameters in the linear relationship yields total molecular energies of -283.06851 and -3151.210613 hartree for the 15-benzene and 16-benzene rings, respectively, which are in excellent agreement with the calculated total molecular energies (see Table1). The *change $\Delta E$ in total molecular energy* between polymers having successive values of *n* for the *ab-initio* case also converges rapidly to a constant value as shown in Table 1 and in Figure 2. *Molecular mechanics* optimizations, using the *universal force field* UFF on the same set of spiral benzene structures also scale*s* linearly with increasing *n* for $n > 3$ and yield the least-squares straight line with parameters

$a$ = -0.0005422 hartree          slope $k$ = 0.027593 hartree/ring

$r^2 = 0.99985$               $\sigma = 0.001496$

From Table 1 and Figures 2 and 3 it is further clear that *extrapolations* for the total molecular energy from $n = 10$ to $n = 15$ and $n = 20$ almost exactly match the values which were obtained by directly calculating the optimized total energies. *Extrapolations to larger values of n can thus be made with confidence*.

The energy difference between the lowest and the highest energies in molecules of the present study covers an enormous molecular energy range of almost 1400 hartree, or 38 keV for the G03-*ab-initio* computations. It is clear that the predictions of the model proposed here adequately describe the variation of the total molecular energy $E_k$ with the number $n$ of benzene rings.

As illustrated in Figures 3 and 4 the assumption inherent in the formation of the molecular graphs of Figure 1, namely that the sum of the short-range and long-range interactions saturates and becomes linear for $n \geq 3$ and 4 respectively is thus supported. Computations with smaller values of $n$ can therefore yield reliable energy values for larger polymer sizes when linearly extrapolated. Similar results were also been obtained for other compounds, such as the *planar* zigzag polybenzenes, which will be published elsewhere [17].

The following conclusions follow from the preceding discussion:

(i) The computations reported above verify the validity of the model assumed, that is, although the contribution of short-range interactions dominate the contributions to the total molecular energy, computational methodologies must be used which include *all effects* which contribute to produce the energy minimum of the optimized geometry of polymers.

(ii) The model predicts a linear scaling of polymer energies with an increasing number of rings, *which is independent of the computational methodology used*.

(iii) The linear scaling determined for a series of small values of $n$ for a particular polymer sequence allows extrapolation to larger values of $n$ which are out of computational reach because of the sheer size of the computational restraints of time, memory and storage space.

(iv) The topological relationships inherent in Figure 1 yield the correct model for the symbolic construction of linear polymers.

The authors thank Pebble Bed Modular Reactor (Pty) Ltd. for supporting the University of Pretoria in the purchase of the IBM cluster described in the text.

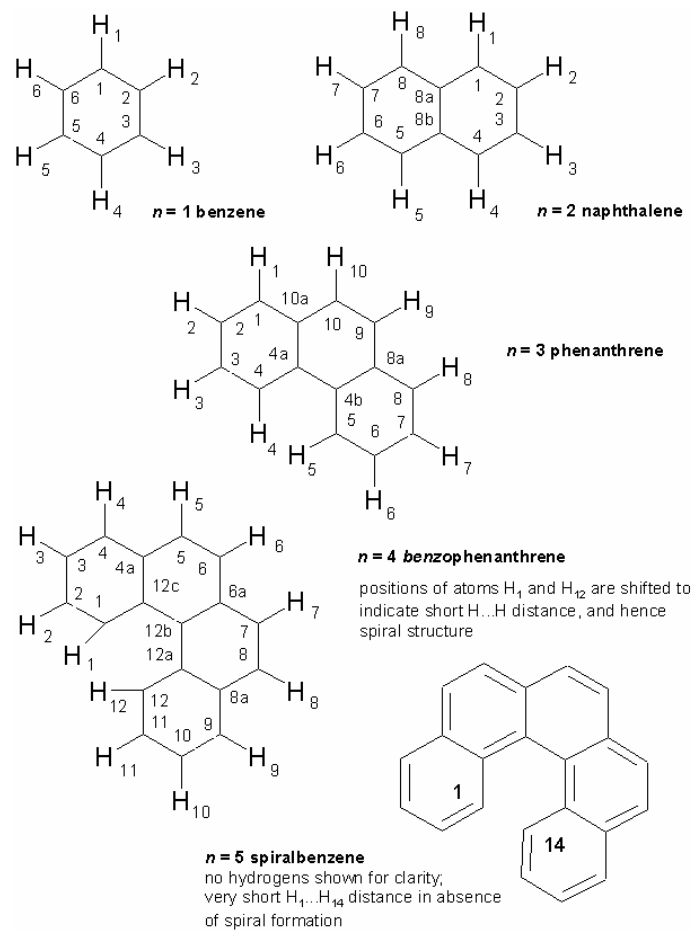

Figure 1. The first few spiralbenzenes (see text), showing IUPAC numbering of atoms.

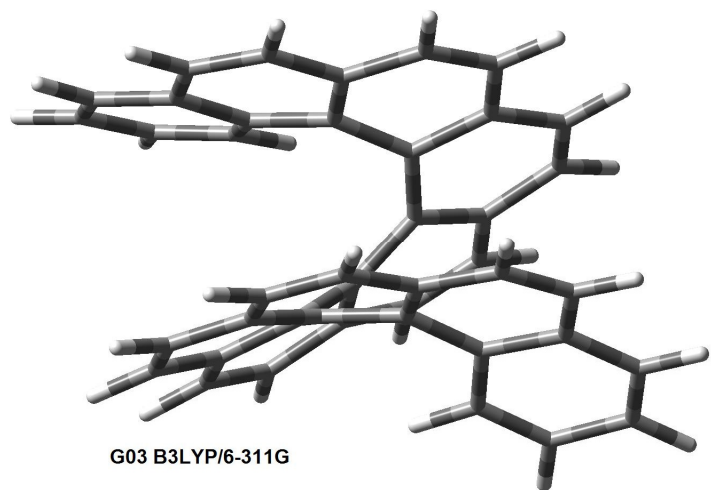

**G03 B3LYP/6-311G**

**Optimized structure of an ortho-fused spiral-benzene with 10 rings**

Figure 2. *Ortho*-addition depicted, with perspective view of the optimized geometrical structure of a typical spiral-benzene polymer containing 10 benzene rings, where *n* = 10.

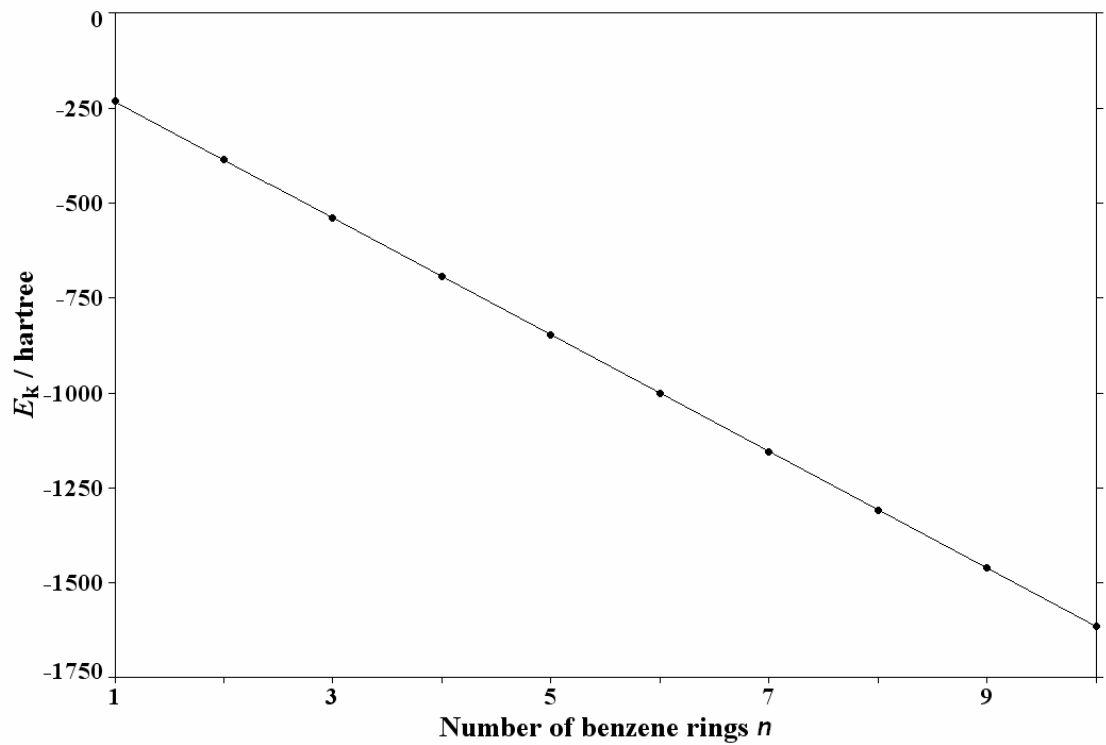

Figure 3. The total molecular energy $E_k$ with the number of benzene rings n as calculated by the ab-initio DFT method described in the text. . The solid curve is a least-squares fit to $E_n = a + kn$ with $a = -0.78.642213$ hartree and $k = -153.62842$ hartree/ring with points for $n = 1, 2, 3$ and 4 excluded in the fit (see text), $r^2 = 0.99985$ (correlation coefficient) and $\sigma = 0.001496$ (standard deviation) are obtained for the fit.

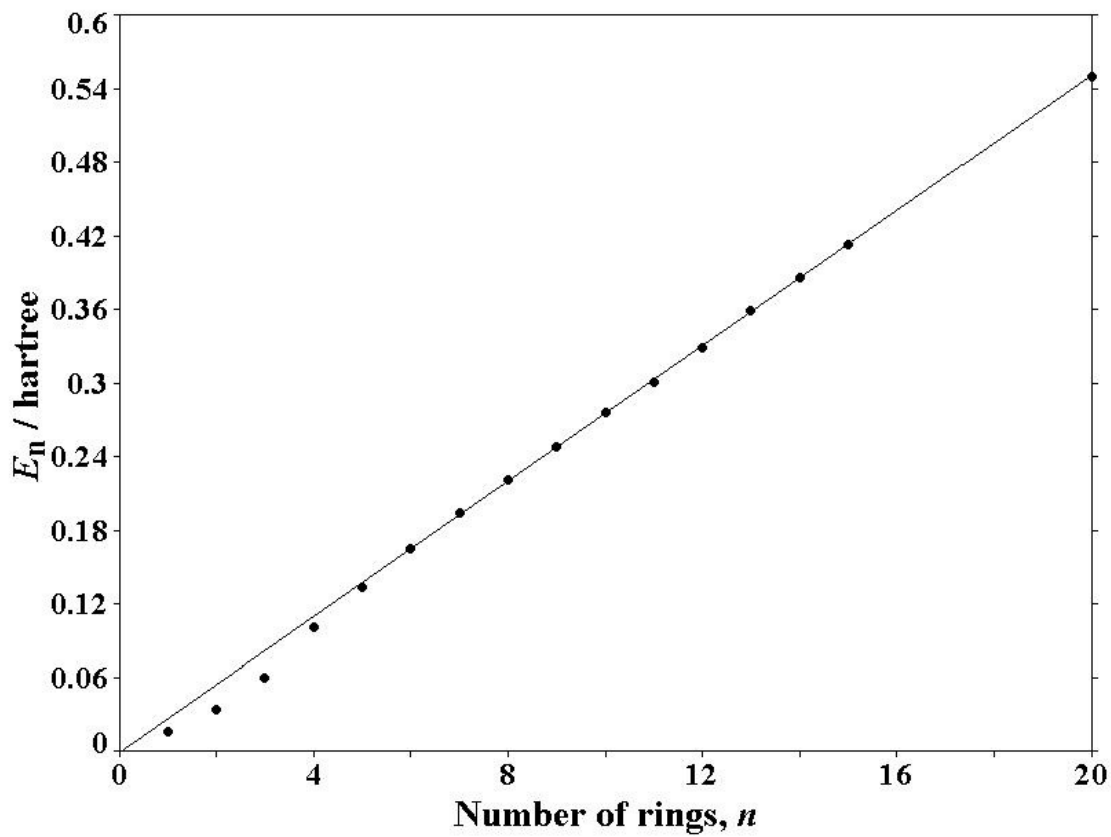

Figure 4. The total MM optimized energies $E_n$ as a function of the number of benzene rings, $n$. The solid curve is a least-squares fit to $E_n = a + kn$ with $a$ = -0.00054219043 hartree and $k$ = 0.027593312 hartree/ring and points for $n$ = 1, 2, 3 and 4 not included in the fit. $r^2$ = 0.99985 (correlation coefficient) and $\sigma$ = 0.001496 (standard deviation) are obtained for the fit.

Table 1. The total molecular energy, $E_n$, calculated using the ab-initio DFT method described in the text for the spiral benzenes in hartree, where $n$ is the number of benzene rings. The energy differences between successive molecules, and the double differences are in hartree.

| $n$-benzene | $E$ | $\Delta E_{n+1} = E_n - E_{n+1}$ | $\Delta(\Delta E)_{n+2} = \Delta E_{n+1} - \Delta E_{n+2}$ |
|---|---|---|---|
| 1-benzene | -232.248144586 | | |
| 2-benzene | -385.886213324 | 153.638068738 | |
| 3-benzene | -539.526475268 | 153.640261944 | -0.002193206 |
| 4-benzene | -693.154127782 | 153.627652514 | 0.01260943 |
| 5-benzene | -846.782985022 | 153.628857240 | -0.001204726 |
| 6-benzene | -1000.41355560 | 153.630570578 | -0.001713338 |
| 7-benzene | -1154.04202930 | 153.628473700 | 0.002096878 |
| 8-benzene | -1307.66995698 | 153.627927680 | 0.000546020 |
| 9-benzene | -1461.29832126 | 153.628364280 | -0.000436600 |
| 10-benzene | -1614.92550793 | 153.627186670 | 0.00117761015 |
| 15-benzene | -2383.06032953 | | |
| 20-benzene | -3151.19198763 | | |

.....................................................................................................